\documentclass[epsf]{aa}
\usepackage{natbib}
\usepackage{epsfig}

\begin{document}
\title{Improved parameters for the transiting planet HD\,17156b: a high-density giant planet with a very eccentric orbit\thanks{Based on observations made with the Mercator Telescope, operated on the island of La Palma by the Flemish Community, at the Spanish Observatorio del Roque de los Muchachos of the Instituto de Astrof\'isica de Canarias.}}
\subtitle{}
\author{M. Gillon$^{1}$, A. H. M. J. Triaud$^1$, M. Mayor$^1$, D. Queloz$^1$, S. Udry$^1$, P. North$^2$}    

\offprints{michael.gillon@obs.unige.ch}
\institute{$^1$  Observatoire de Gen\`eve, Universit\'e de Gen\`eve, 51 Chemin des Maillettes, 1290 Sauverny, Switzerland\\
$^2$ Laboratoire d'Astrophysique, Ecole Polytechnique F\'ed\'erale de Lausanne (EPFL), Observatoire de Sauverny, CH-1290 Versoix, Switzerland\\}

\date{Received date / accepted date}
\authorrunning{M. Gillon et al.}
\titlerunning{Improved parameters for the transiting planet HD 17156b}
\abstract{We report  high-precision transit photometry for the recently detected planet HD\,17156b. Using these new data with previously published transit photometry and radial velocity measurements, we perform a combined analysis based on a Markov Chain Monte Carlo approach. The resulting mass $M_p = 3.09$ $^{+0.22}_{-0.17}$ $M_{Jup}$ and radius $R_p =  1.23$ $^{+0.17}_{-0.20}$ $R_{Jup}$ for the planet places it at the outer edge of the density distribution of known transiting planets with $\rho_p = $ 1.66 $^{+1.37}_{-0.60}$ $\rho_{Jup}$. The obtained transit ephemeris is $T_{Tr} = 2454438.48271$ $^{+0.00077}_{-0.00057} + N \times 21.21747$ $^{+0.00070}_{-0.00067}$ BJD. The derived plausible tidal circularization time scales for HD\,17156b are larger than the age of the host star. The measured high orbital eccentricity $e = 0.6719$ $^{+0.0052}_{-0.0063}$ can thus not be interpreted as the clear sign of the presence of another body in the system. 
 }

\maketitle

\section{Introduction}

Currently, exoplanets that transit their parent stars are undoubtedly the most important source of information about the physics and composition of the planetary objects outside our Solar System (see review by Charbonneau et al. 2007). The discovery rate of such transiting planets has increased recently thanks mainly to the excellent efficiency of ground-based wide-field surveys such as WASP (Pollaco et al. 2006) and HAT (Bakos et al. 2002). Also, a thorough characterization of the few transiting planets orbiting stars that are bright enough has brought very interesting results, most  due to the high capabilites of the Spitzer Space Telescope (see e.g. Harrington et al. 2007, Knutson et al. 2007). With the space mission CoRoT that is now in operation (Baglin et al. 2006) and the future launch of Kepler (Borucki et al. 2007) and JWST (Gardner et  al. 2007), we can expect that transiting planets will continue to play a major role in our understanding of extrasolar planets in the coming years.

Most of the known transiting planets are hot Jupiters, i.e. very short period (less than 5 days) tidally circularized planets with masses ranging from $\sim 0.5$ to $\sim$ 2 $M_{Jup}$ and densities spanning a rather large range with an upper limit close to that of Jupiter. Nevertheless, some planets very different from this description have been observed recently in transit. Among them are the very massive HD\,147506b (Bakos et al. 2007), CoRoT-Exo-2b (Alonso et al. 2008)  and XO-3b (Johns-Krull et al. 2008), and also the hot Neptune GJ\,436b (Butler et al. 2004; Gillon et al. 2007a). Interestingly, three of these four planets have a non-null eccentricity despite their small periods.

Another exceptional transiting planet was announced recently: HD\,17156b (Fisher et al. 2007; Barbieri et al. 2007). It orbits around a bright (B=8.8,V=8.2) G0V star. Its period $P \sim$ 21.2 days is by far the longest one among the transiting planets. Furthermore, this massive planet ($M  \sim  3.1 M_{Jup}$) has a very eccentric orbit ($e \sim$ 0.67). After GJ\,436b, it is the second one for which the transiting status is detected after the announcement of the radial velocities (RV) orbit. This transit detection (Barbieri et al. 2007) was done under the auspices of the {\tt TransitSearch.org} network (see e.g. Shankland et al. 2006) which is based on a collaboration between professional and amateur astronomers and aims to detect the possible transits of the planets detected by RV. 

One transit of HD\,17156b was observed by Barbieri et al. (2007). The quality of their photometry was high enough to detect the transit with a good level of confidence. Nevertheless, obtaining a more precise transit lightcurve at a different epoch was desirable to constrain more thoroughly the transit parameters and  to obtain a more precise orbital period than the one deduced from RV measurements ($21.2 \pm 0.3$ days). This motivated us to observe another transit of HD\,17156b on December 3th 2007 from La Palma with the 1.2m Mercator Belgian telescope. We present these observations in Sec. 2, and their reduction is described in Sec. 3. We analyzed this new photometry in combination with published transit photometry and RV measurements using a method based on a Markov Chain Monte Carlo (MCMC) approach described in Sec. 4. The results of our analysis are presented in Sec. 5 and discussed in Sec. 6. 

\section{Observations}

Based on the ephemeris presented in Barbieri et al. (2007), a transit of HD\,17156b was expected to be clearly visible from Canary Islands during the night of December 3, 2007. We observed it with the 1.2m Mercator Belgian telescope located at the Roque de los Muchachos Observatory on La Palma Island. The instrument used was the MEROPE CCD camera. It has a field of view (FOV) of 6.5' by 6.5' and a pixel scale of 0.19''. A set of 213 exposures were taken in the B2 filter ($\lambda_{eff}= 447.8$ nm, $\Delta\lambda $ = 13.9 nm) from 19h54 to 04h34 UT. The exposure time varied from 30s to 60s. A large defocus was applied to obtain a good trade-off between duty cycle,  time sampling and scintillation mitigation. Transparency conditions during the night were good. The airmass decreased from 1.57 to 1.37 then increase to 2 at the end of the run.

During the first out-of-transit (OOT) part of the run, a problem of defocus adjustment caused a minority of the pixels of HD 17156b image to fall outside the linearity range for parts of the images. This problem was fixed just before the ingress. Another technical problem occured in the dome during the transit that led to a loss of $\sim 30$ min of observation. Fortunately, this problem occured during the long bottom of the transit, not in the ingress or egress. 

\section{Data reduction}

After a standard pre-reduction, all images were reduced with the IRAF/DAOPHOT aperture photometry software (Stetson, 1987). As the defocus was not the same for the whole run, the reduction parameters were adapted to the FWHM of each image. Differential photometry was then performed using the flux of the nearby star BD+71 168 (B=V=9.6) as the reference flux. The resulting lightcurve was finally decorrelated for airmass variations using its OOT parts\footnote{Our final photometric time series is available only in electronic form at the CDS via anonymous ftp to cdsarc.u-strasbg.fr (130.79.128.5)
or via http://cdsweb.u-strasbg.fr/cgi-bin/qcat?J/A+A/. }. No correlation with the other external parameters was found.

The $rms$ for the first OOT part is $2\times10^{-3}$. It is $\sim$ 2 times the theoretical error bar per point. As can be seen clearly in Fig. 1, this part of the curve is noisier and less populated than the rest of the curve. The cause is the too small defocus in the first part of the run and the resulting linearity problems. The rest of the curve is better. The $rms$ of the residuals of the fit during the transit is $1.2\times10^{-3}$, while it is  $1.7\times10^{-3}$ for the second OOT part. The increase of the noise at the end of the run is due to the increase of the airmass and the resulting increase of the scintillation. 

We estimated the level of red noise $\sigma_r$ in our photometry using the equation (Gillon et al. 2006):

\begin{equation}\label{eq:g}
\sigma_r =  \bigg(\frac{N\sigma_N^2 - \sigma^2}{N - 1}\bigg)^{1/2}\textrm{,}
\end{equation}

\noindent
where $\sigma$ is the $rms$ in the original OOT data and $\sigma_N$ is the standard deviation after binning the OOT data into groups of $N$ points. We used $N = 10$, corresponding to a bin duration similar to the ingress/egress timescale. The obtained value for $\sigma_r$ is compatible with purely Gaussian noise.


\begin{figure}
\label{fig:a}
\centering                     
\includegraphics[width=9cm]{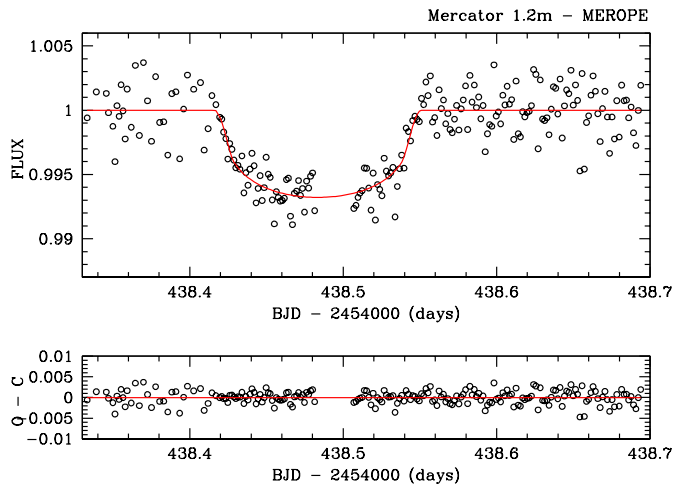}
\caption{$Top$: Mercator/MENOPE photometry for the transit of HD\,17156b. The best fitting theoretical transit curve is superimposed in red. $Bottom$: residuals of the fit ($rms$=1.6e-3).}
\end{figure}

\section{Data analysis}

In addition to our Mercator photometry, we used the `Almenara'  transit photometry (Barbieri et al. 2007) and published Keck and Subaru RVs (Fischer et al. 2007) to determine the parameters of the system. The data were analysed with a program called FullTransit. It carries out a combined multi-band photometry and spectroscopy fit based on a MCMC approach. The models used for the photometry are, that of Mandel \& Agol (2002) and Gim\'enez (2006a) with a quadratic lim darkening law, for the spectroscopy, a standard eccentric orbit model, and for the Rossiter-McLaughlin effect, Gim\'enez (2006b) models. Since  no spectroscopic measurements were taken during transit, the Rossiter effect is presented for information purposes.

\subsection{Motivations}

It seems normal to use all the available data  in order to characterise a planet. By fitting at the same time different models in various datasets, the fit finds the best compromise between all of them and also  reveals discrepancies between timings if they arise. It is also a way of remaining consistent in the model fitting of the data.

Our choice of MCMC is due to the now abundant literature on the subject (see Collier Cameron et al. 2007 and references thererein), but also because it seems the easiest way to have a single set of parameters common to each model for the fit. Also, the MCMC allows us to determine precise errors on the model from the simulations performed.

\subsection{Method}

The program currently fits 10 free parameters: the depth of the transit $D$,  the rotational velocity of the star $V sin I$, the RV semi-amplitude $K$, the impact parameter $b$, the width of the transit $W$, the angle between the equatorial plane of the star and the orbital plane of the planet $\beta$, the orbital period $P$,  the middle of transit date $T_{tr}$, the eccentricity $e$, and the angle to the periastron $\omega_0$. Most of these parameters are directly observable or can be estimated from the data. This set has been choosen so as to minimize the correlation between parameters, which could  lead to a non convergence or a biased result. 

From these 10 parameters, physical parameters needed by the models are calculated. The model is created in phase for each point and $\chi^2$ statistics is used to estimate the goodness of fit. A penalty on the $\chi^2$ using Bayesian errors can be added if necessary.

In addition to these 10 parameters, there is also an optimal scaling for each set of data. In the current situation, we have 2 sets of radial velocity and 2 sets of photometry, hence 4 other parameters. These do not participate in the MCMC, they are rather the result of the $\chi^2$ statistics.

Each of the parameters is calculated as follow:
\begin{eqnarray}
parameter_2(j) = parameter_1(j)\,\sigma(j)\,G(0,1)\,f
\end{eqnarray}
\noindent
where $\sigma(j)$ is the standard deviation of the parameter, $G(0,1)$ is a random Gaussian number, and $f$ is a factor ensuring that 25\% of the MCMC steps are being accepted. For each step in the MCMC a set of 10 parameters is created. This step is chosen to be accepted or not by a Metropolis-Hastings algorithm (see  Collier Cameron et al. 2007).

At the start of the MCMC, some guessed parameters are inserted along with their respective $\sigma$s. These have been fixed to make sure that each parameter explores randomly the parameter space around their best $\chi2$. The $\sigma$s act as the error on the prior; $f\sigma(j)$ is the step size of the MCMC, f is estimated every 100 steps to make sure that 25\% of the steps are accepted.

 After $n$ steps, the best $\chi^2$ is found and its associated set of parameters becomes the best fit. The other sets are scrutinised and the 68.3\% sets around the best fit give the error. It is not calculated using a $\Delta\,\chi^2$ because some of the distributions are not Gaussian.

It has been decided not to have limb darkening coefficients as free parameters so  to not overload the MCMC as well as to avoid discrepancies in stellar parameters between individual photometric bands and a fitting of the Rossiter.  Using Mandel \& Agol (2002), this would add 4 more parameters, 4 in the case of fitting using the Gim\'enez models, plus 2 others used for the Rossiter. As FullTransit is a characterization  program, previoulsy independently fitted limb darkening coefficients can be inserted in the calculations of the models.

For a convergence, it is necessary to have a good idea of the period. We thus first fitted a theoretical transit on the Mercator photometry using the method described in Gillon et al (2007b), then used the obtained timing and the one presented in Barbieri et al. (2007) to deduce a precise initial guess for the orbital period before starting the MCMC. 

FullTransit is used to characterize the parameters of a planet, not to find them. Once the period is found, it is  straightforward to find parameters approaching the best fit and to launch the MCMC. These, though, should not be too close to the final solution - if known in advance - so as to let the MCMC explore the $\chi^2$ potential around the solution. 

\subsection{Analysis}

A  $2$ $m.s^{-1}$ error was added to the existing error for the radial velocity measurements to allow for the jitter reported in Fischer et al. (2007). 

The limb darkening coefficients were extracted for the two bands $B2$ and $R$ from the table produced by Claret (2000) for the quadratic law.
These were selected for a $6000\,K$ star with $log\,g = 4.5$, [Fe/H] = 0.2, close to the physical parameters presented in Fischer et al. (2007). A stellar mass of $1.2 \pm 0.1 M_{\odot}$ was used (Fischer et al. 2007); the mass was inserted randomly as $M_{\star} = 1.2+0.1\,G(0,1)$ and a Bayesian error was added to $\chi^2$ to make a quality function $Q_j = \chi^2_j + {(1.2-M_{\star,j})^2 \over{0.1^2}}$ at each step $j$.

Two analyses could be done, one using the Mandel \& Agol (2002) models, the other Gim\'enez (2006a). Both were performed and being similar, only the Mandel \& Agol (2002) was pursued, as it took much less time to run than the other. If there has been a radial velocity point during the transit, the whole analysis could have been conducted using Gim\'enez (2006a and b) to remain consistent throughout the fitting process. For the same reason, because there is no Rossiter involved with this star, the two parameters $V\,sin\,I$ and $\beta$ are not effectively used in the fit. 

Various starting parameters were tried on a range larger than the error bars on the final parameters calculated by each chain. All chains converged to within the error bars of each other. The chains allowed a large safety burn-in period of 15\,000 steps, and a simulation of 100\,000 steps each. The final results give an average over the chains that were calculated. The probability of a secondary transit was estimated by examining how many sets of parameters have a secondary eclipse impact parameter $b_{sec}< 1-{R_p\over R_{\star}}$. For a grazing secondary transit, the probability is not much higher and would probably not be observable.

\section{Results}

\begin{table}
\begin{tabular}{ll} \hline \hline 
{\it MCMC fitted parameters}  & \\
$D$                           &  $0.00605$ $^{+0.00041}_{-0.00051}$      \\
$K$ [m.s$^{-1}$]     &  $272.6$ $^{+ 4.5}_{-4.2}$ \\
$b$                            & $0.591$ $^{+ 0.088}_{-0.191}$      \\
$W$ [phase]            & $0.00626$ $^{+ 0.00015}_{-0.00016}$      \\
$P$ [days]                & $21.21747$ $^{+ 0.00070}_{-0.00067}$    \\
$T_{tr}$ [BJD]            & $2454438.48271$ $^{+0.00077}_{-0.00057}$      \\
$e$                            & 0.6719 $^{+ 0.0052}_{-0.0063}$       \\
$\omega_0$ [$^o$] & $121.14$ $^{+ 0.76}_{-0.89}$\\ 
& \\
{\it Deduced transit parameters}  & \\
$p$ = $R_p/R_{\star}$        & $0.0777$ $^{+ 0.0026}_{-0.0034}$\\
$r_{\star}$ = $R_{\star} /a$  & $0.0476$ $^{+ 0.0045}_{-0.0058}$\\
$r_{p}$ = $R_{p} /a$           & $0.00371$ $^{+ 0.00047}_{-0.00059}$\\
& \\
{\it Deduced stellar radius$^\ast$} & \\
$R_{\star}$ [$R_\odot$] &  1.63 $^{+ 0.17}_{-0.20}$\\
& \\
{\it Deduced planetary parameters}  & \\
$M_{p}$ [$M_{Jup}$]& $3.09$ $^{+ 0.22}_{-0.17}$ \\
$R_{p}$ [$R_{Jup}$]&  $1.23$ $ ^{+ 0.17}_{-0.20}$ \\
$\rho_p$[$\rho_{Jup}$]&   $1.66$ $^{+1.37}_{-0.60}$ \\
& \\
{\it Deduced orbital parameters}&\\
a [AU] & $0.1589$ $^{+ 0.0054}_{-0.0044}$\\
$i$ [$^o$] & $85.4$ $^{+ 1.9}_{-1.2}$\\
\\
Probability of secondary eclipse & $0.04039 \%$\\
\hline 
& \\
\end{tabular}
\caption{Fitted and derived parameters for the HD\,17156 system, host star and transiting planet. See Sec. 4.2. for a description of the fitted parameters. $^\ast$: the value for the stellar mass was kept fixed to the one presented in Fischer et al. (2007): $1.2 \pm 0.1 M_\odot$.} 
\label{param}
\end{table}

Table 1 shows the 8 fitted parameters. For each set of parameters recorded by the simulations, a set of physical parameters - deduced transit parameters and stellar, planetary \& orbital parameters - was calculated. The error on physical parameters was calculated the same way as the fitted parameters: by taking 68.3 \% of the sample around the best fit.

The impact parameter is the most volatile parameter in this fit, because of a lack of data at the bottom of the lightcurve trough.

An average $Q$ was estimated as 684.57 (25.99 on the spectroscopic data, 658.55 on the photometric data) giving an overal reduced $Q = 1.63$.

Fig. 2 shows the global fit of the data, including the Rossiter-McLaughlin effect as it would occur for a $V\,sin\,I$ of 2.8 $km.s^{-1}$. Its amplitude is small due to the high slope of the eccentric orbit.

\begin{figure}
\label{fig:b}
\centering                     
\includegraphics[width=9cm]{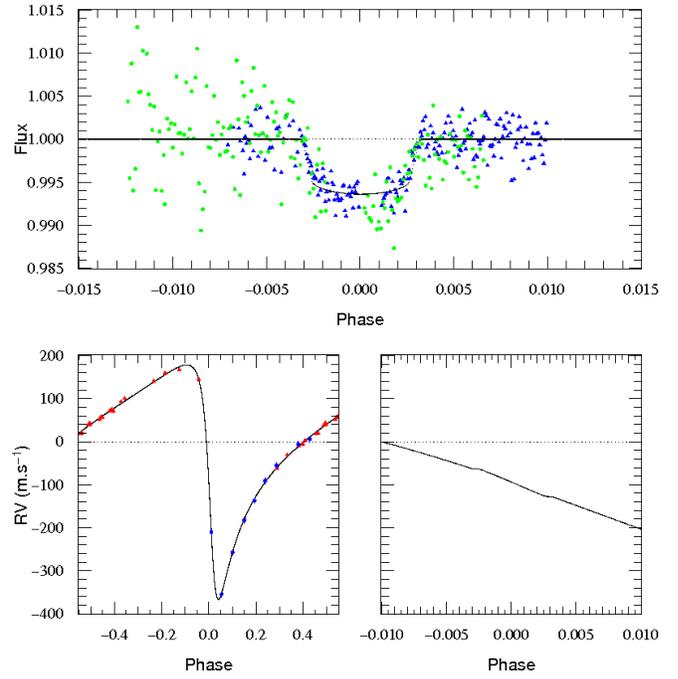}
\caption{Model fitted using FullTransit on the photometric ($top$) and RV ($bottom$) data. $Bottom$ $left$: red triangles = Keck, blue circles = Subaru. $Bottom$ $right$: zoom on the transit phase. $Top$: blue triangles = Mercator photometry, green circles = `Almenara' photometry.}
\end{figure}

Optimal scaling for the Keck data is 93.8241 $m.s^{-1}$ while for the Subaru it is 93.1092 $m.s^{-1}$. The optimal scaling for the Mercator photometry is a factor of 1.00005, while for the Almenara data, it is 1.0018.

\section{Discussion}




Irwin et al. (2008) presented new  photometry for one transit of HD\,17156b observed from three separate observatories. The agreement between their deduced parameters and ours is satisfactory. Our error bars on the planet radius are larger than theirs, and this comes from the fact that our photometry was good enough to independently determine the stellar radius, while Irwin et al.  applied a Bayesian constraint on the stellar radius to keep its fitted value close to the one determined by spectroscopic analysis (Fischer et al. 2007). We notice that our error bars on the planet and star radius are larger than 10\%, so the characterization of this system would benefit from further high-precision transit photometry.

Fortney et al. (2007) presented theoretical radius values for planets over a wide range of masses, using  realistic atmospheric boundary conditions and equations of state for core materials. For a planet similar to HD\,17156b, their theoretical radii range from 1.02 $R_{Jup}$ for a 100 $M_\oplus$ core  to 1.1 $R_{Jup}$ if no core is present. The presence of a core has thus a very weak influence on the planetary radius in this planetary mass regime. Our measured radius seems to argue against a heavy core for HD\,17156b, but the error bar is too large to constraint the core mass. As can be seen in Fig. 3, HD\,17156b has the highest density among all transiting exoplanets except HD\,147506b and XO-3b.

\begin{figure}
\label{fig:a}
\centering                     
\includegraphics[width=9cm]{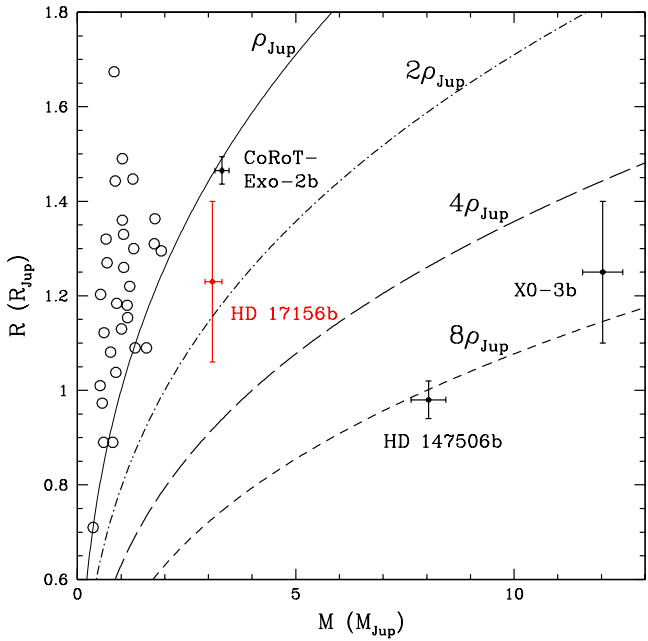}
\caption{Mass radius diagram for the known transiting planets with an error bar on the mass and radius smaller than 10 \% (except GJ\,436b). While most of them ($open$  $circles$) have a density comparable to or lower than the one of Jupiter , the three massive planets HD\,17156b, HD\,147506b and XO-3b are much denser, as predicted by theory, while CoRoT-Exo-2b  appears to be `anomalously' large. The 1-sigma error bars are represented only for the 4 massive planets for clarity. }
\end{figure}

The very high orbital eccentricity of HD\,17156b could indicate intense interaction with a still undetected third body. The circularization timescale $\tau_{circ}$ of a planet can be computed using (Goldreich \& Sotter 1966): 

\begin{equation}\label{eq:i}
\tau_{circ} = \bigg( \frac{2 Q_p P}{63 \pi} \bigg)  \bigg( \frac{M_p}{M_{\ast}} \bigg)  \bigg( \frac{a}{R_p} \bigg)^5 
\end{equation}
\noindent
where $Q_p$ is the tidal quality factor.  For Jupiter, $Q_p$ is estimated to lie between $10^5$ and $2 \times 10^6$ (Goldreich \& Soter, 1966; Peale \& Greenberg 1980) Assuming the lowest of these values for the tidal quality factor of HD 17156b, the obtained value for $\tau_{circ}$ is 208 Gyr, largely exceeding the estimated stellar age of 5.7 Gyr (Fischer et al. 2007). Thus, the high eccentricity of HD\,17156b does not necessarily  indicate the presence of a third body in the system. If nevertheless such a perturber is present, it could be detected with more RV measurements or through the precise measurement of a large number of transits of HD\,17156b, as was attempted for several other transiting planets by e.g. the TLC project (Holman \& Winn 2006).  In the case of a mean-motion resonance, the amplitude of the timing variations is proportional to the period of the perturbed body (see e.g. Holman \& Murray 2005). As HD\,17156b has the longest period among the known transiting planets, a dedicated monitoring of its transits should thus have a good sensitivity to any other planet in resonance with it.

HD\,17156b is very different from all the other known transiting planets, and measuring its thermal emission would be very desirable to study its atmospheric heat distribution efficiency, albedo and chemical composition, but  the secondary eclipse probability we obtain is unfortunately very near to zero (0.04\%).

\begin{acknowledgements} 
The authors thank F. Courbin and G. Meylan for making possible the observations with the Belgian Mercator telescope. J. M. Almenara, R. Alonso and M. Barbieri are gratefully acknowledged for providing us with their photometric data. A. H. M. J. Triaud thanks A. Collier Cameron and A. Gim\'enez for all their help and support. This work was supported by the Swiss National Fund for Scientific Research
\end{acknowledgements} 

\bibliographystyle{aa}
{}
\end{document}